\documentclass[10pt,journal]{IEEEtran}
\usepackage{amsmath}
\usepackage{amsfonts}
\usepackage{graphicx}
\usepackage{float}
\usepackage{tabularx}
\usepackage{amssymb}
\usepackage{pdflscape}  
\usepackage{setspace} 
\usepackage{float} 
\usepackage{color}      
\usepackage{graphicx}        
\usepackage{graphics} 
\usepackage{subfig}                             
\usepackage{epsfig} 
\usepackage{times} 
\usepackage[hyphens]{url}  
\usepackage{comment}
\usepackage{algorithmicx,algorithm}
\usepackage{bbm}
\usepackage{algorithm} 
\usepackage{algpseudocode} 
\usepackage{algorithmicx}
\usepackage{pifont}
\usepackage{cite}
\usepackage{multirow}

\title{\vspace{-0.75em}\text{\small Accepted in the 57th North American Power Symposium, Hartford, Connecticut, 2025 (NAPS 2025)} \\
A Learning-based Hybrid System Approach for Detecting Contingencies in Distribution Grids with Inverter-Based Resources
\thanks{This work is supported in part by National Science Foundation under Grants DMS-2229109.\\
* Corresponding Author}}

\author{
    \begin{tabular}{c@{\hspace{4cm}}c} 
       \textbf{Hamid Varmazyari} & \textbf{Masoud H. Nazari*} \\
       \textit{Electrical and Computer Engineering} & \textit{Electrical and Computer Engineering} \\
      \textit{Wayne State University} & \textit{Wayne State University} \\
       Detroit, USA & Detroit, USA \\
       \text{varmazyari.h@wayne.edu} & \text{masoud.nazari@wayne.edu}
    \end{tabular}
  \vspace{-3em}
}

\begin{document}
\maketitle
\begin{abstract} 
This paper presents a machine-learning based Stochastic Hybrid System (SHS) modeling framework to detect contingencies in active distribution networks populated with inverter-based resources (IBRs). In particular, this framework allows detecting unobservable contingencies, which cannot be identified by normal sensing systems. First, a state-space SHS model combining conventional and IRB-based resources is introduced to formulate the dynamic interaction between continuous states of distribution networks and discrete contingency events. This model forms a randomly switching system, where parameters or network topology can change due to contingencies.  We consider two contingency classes: (i) physical events, such as line outages, and (ii) measurement anomalies caused by sensor faults. Leveraging multivariate time series data derived from high-frequency sampling of system states and network outputs, a time series-based learning model is trained for real-time contingency detection and classification. Simulation studies, carried out on the IEEE 33-bus distribution system, demonstrate a 96\% overall detection accuracy. 

\end{abstract}

\section{Introduction}
\label{Sec1}
Power systems are critical infrastructures that ensure the continuous and reliable supply of electricity, forming the backbone of modern societies \cite{ABDELKADER2024102647}. Traditionally, these systems have been dominated by synchronous generators (SGs), which inherently provide system stability through inertia and well-defined dynamic responses \cite{JAFARI2024110067}. However, the evolving landscape of energy generation has led to an increasing integration of inverter-based resources (IBRs), such as solar photovoltaic systems (PV), wind turbines, and battery energy storage systems (BESS). While IBRs offer environmental and operational advantages, they present unique challenges for grid stability due to their lack of inherent inertia and distinct dynamic behaviors \cite{AHMED2023486,biswas2025resilience}.

Given these developments, power system contingencies— power line outages, generator failures, or system faults—pose significant risks to grid reliability \cite{8700246}. The presence of IBRs can exacerbate these risks, as their fast and complex dynamics may lead to rapid system destabilization if contingencies are not detected and mitigated promptly. Therefore, the timely and accurate detection of contingencies in modern power grids is crucial to prevent cascading failures and ensure continuous power delivery \cite{che2019identification}. Addressing these challenges requires advanced detection strategies that account for the unique characteristics of IBRs and their interactions with traditional grid components.

 Contingencies in power systems can be effectively modeled as stochastic jumps within system dynamics, represented as discrete events \cite{10003976}. On the other hand, power system dynamics consist of continuous physical variables, such as voltage and frequency, and discrete variables that characterize sudden contingencies like faults or equipment failures. This dual nature of system behavior naturally lends itself to a Stochastic Hybrid System (SHS) framework, which captures the interaction between continuous evolution and discrete stochastic transitions. By adopting the SHS model, it becomes possible to more accurately represent and analyze the complex dynamics of power systems, particularly in the presence of IBRs\cite{YUAN202410141,MehdipourAbadi2025}. This modeling approach enhances the understanding of how random, discrete events influence the continuous state of the system, thereby supporting more effective contingency detection and mitigation strategies. 
 
 Various methods have been employed for contingency detection in power systems\cite{YUAN202410141,en14040923,ZAREGHALEHSEYYEDI2023108670,9763039,abdolahi2021reliability,abdelmalak2022enhancing,9664189}. Traditional approaches rely on model-based techniques such as state estimation, stability analysis, and fault current measurements \cite{ZAREGHALEHSEYYEDI2023108670,9763039,abdolahi2021reliability}. While effective for conventional synchronous generator systems, these methods often struggle to adapt to the dynamic and less predictable nature of IBRs. Moreover, some research efforts focus solely on either traditional systems or IBR-specific models, neglecting the complexities introduced by their integration \cite{YUAN202410141,en14040923,ZAREGHALEHSEYYEDI2023108670,9763039,abdolahi2021reliability}. In contrast, this paper proposes a combined analytical model that captures the interactions between traditional and inverter-based resources, providing a more comprehensive framework for contingency detection.

In recent years, data-driven learning methods have gained attention, leveraging historical data for contingency detection without relying on analytical models \cite{yu2024interpretable,lagos2021data,golpira2022data,liu2021data}. While these methods offer flexibility, they often lack the interpretability and robustness required for complex system dynamics. To address these limitations, this paper integrates an analytical SHS model with advanced time series learning methods, leveraging their strength in capturing temporal patterns and dynamic behaviors. Time series learning methods offer significant advantages over traditional static approaches by enhancing prediction accuracy and adaptability to evolving system conditions \cite{Kong2025,10870219}. This combined approach aims to enhance contingency detection capabilities in modern power systems, ensuring greater reliability and stability.

This paper presents a time series learning-based framework for contingency detection in distribution grids with high penetration of IBRs. The main contributions of this work are as follows: (1) the development of a novel SHS model that accurately represents the dynamic interactions between conventional and IBR power systems, capturing both continuous and discrete system behaviors; and (2) the introduction of advanced learning methods for contingency detection. 
By integrating analytical modeling with cutting-edge time series learning techniques, this framework aims to significantly enhance the accuracy and robustness of contingency detection in complex power system environments.

The rest of the paper is organized as follows:
Section~\ref{Sec2} introduces the integrated dynamic model of the PV–BESS and SGs subsystems.
Section~\ref{sec3} presents the contingency modeling within the SHS framework for distribution grids with IBRs, distinguishing physical and measurement events through a switched system formulation.
Section~\ref{Sec4} presents a Transformer-based time series learning model for real-time contingency detection within the SHS framework, detailing both the offline training and online classification process.
Section~\ref{Sec5} evaluates the proposed methodology through simulations conducted on the IEEE 33-bus distribution system under various contingency scenarios.
Finally, Section~\ref{Sec6} summarizes the key
findings and conclusions of the study.

\section{Integrated State-Space Model of Synchronous Generator and PV–BESS}
\label{Sec2}
In this section, we present the complete system model, encompassing SGs, PV units, and BESS. 
%
We adopt a hybrid system-based modeling framework in which power system buses are classified as either \textit{dynamic} or \textit{non-dynamic}, regardless of their physical configurations \cite{YUAN202410141}. This abstraction departs from the traditional PV/PQ/slack bus classification and enables a unified treatment of distribution grids with high penetration of inverter-based resources, controllable loads, and local controllers. Dynamic buses are modeled using differential equations that capture local dynamics---such as generator swing equations or inverter control loops---while non-dynamic buses are governed by algebraic power balance constraints \cite{YUAN202410141, Glover2001}. For SGs, the dynamics are captured by the classical swing equation
\begin{equation}
M_i \dot{\omega}_i + D_i \omega_i = P_{\text{in}}^i - P_{\text{L}}^i - P_{\text{out}}^i,
\end{equation}
where $M_i$ is the inertia constant, $D_i$ is the damping coefficient, $\omega_i$ is the frequency deviation, and the power terms represent real power input, local load, and power transferred to neighboring buses, respectively. By solving the algebraic equations associated with non-dynamic buses, we derive a reduced-order model that evolves only over the dynamic states. 

\begin{figure}[t!]
  \includegraphics[width=1.5\linewidth, trim={7cm 5cm 10cm 7.5cm},clip]{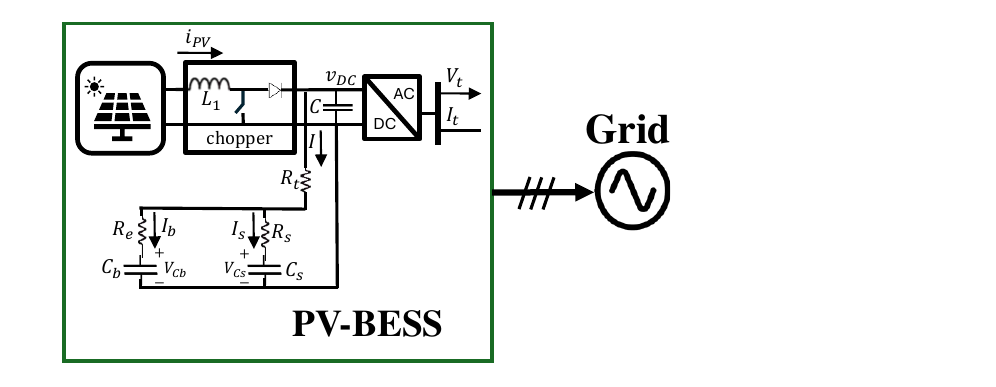}
        \centering
        \captionsetup{width=\linewidth}
        \vspace{-2em}
        \caption{Schematic representation of the PV–BESS system model \cite{MehdipourAbadi2025}. }
        \label{Fig_PV}
        \vspace{-1.5em}
\end{figure} 
\begin{figure}[t]
    \centering
    
    \subfloat[]{\includegraphics[width=1\linewidth]{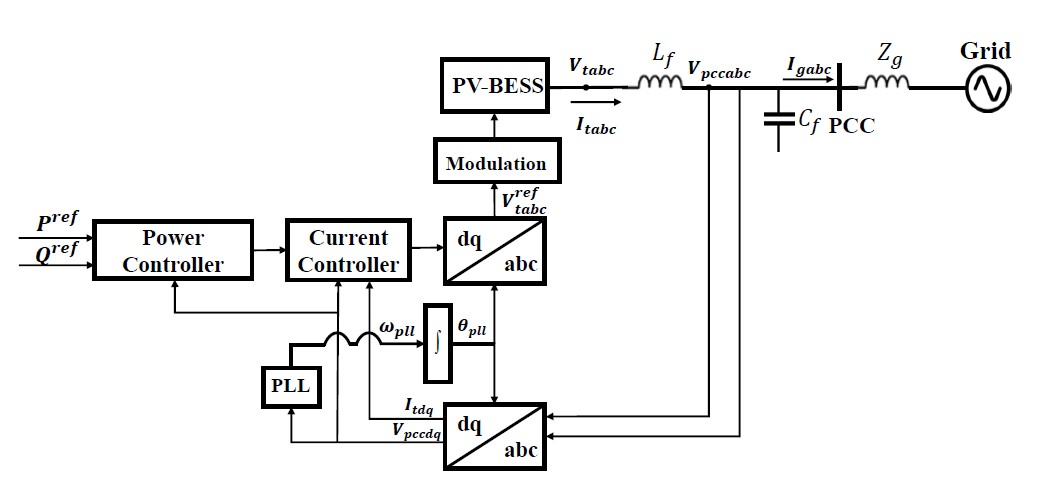}}
    
    \caption{Controller diagrams of the PV–BESS in GFL modes.}
    \vspace{-1.5em}
    \label{fig:control}
\end{figure}

\subsection{PV--BESS Modeling}
\label{ssec:PVBESS_model}


The PV–BESS system comprises a photovoltaic array interfaced via a DC-DC boost converter to a shared DC-link, and a battery storage system represented by a nonlinear equivalent circuit connected in parallel to the same DC-link \cite{MehdipourAbadi2025} (Fig.~\ref{Fig_PV}). Both resources are integrated through a common DC-AC inverter, linking them to the grid through an $L_f$–$C_f$ filter at the point of common coupling (PCC). In this study, a linearized model of the PV–BESS system is employed to facilitate analysis and control design. 

Two primary strategies can be utilized to connect the PV–BESS unit to the grid \cite{electronics13101958}: grid-forming (GFM) and grid-following (GFL). 
%
The GFL mode as shown in Fig. \ref{fig:control}, synchronizes the inverter with the grid voltage via a phase-locked loop (PLL). The PLL generates a synchronization angle $\theta_{pll}$ by measuring and tracking the voltage at the PCC. An external power controller receives active and reactive power references ($P^{ref}$, $Q^{ref}$), generating corresponding dq-axis current references ($i_{tdq}^{ref}$). These current references are then tracked by an inner current controller that regulates the inverter output currents. Afterward, the inverter converts the controlled dq-axis currents to abc coordinates, and the modulation stage generates the required gating signals to regulate the injected currents. Thus, the GFL-controlled inverter operates as a controlled current source. 
%
In this study, we focus specifically on the GFL control mode. 
\subsection{Integrated State-Space Model of Conventional and IRB-based Resources}
This section presents a state-space representation of the integrated system comprising GFL-controlled PV--BESS units and SGs. The dynamics of the GFL-controlled PV--BESS, as shown in Fig.~\ref{fig:control}, are organized into well-defined state subsets \cite{10398460}. Specifically, the state vector is structured as $\Delta x_{GFL} = [\Delta x_{PLL}, \Delta x_{PQ}, \Delta x_{c}, \Delta x_{LCL}, \Delta x_{int}]^T$, where $\Delta x_{PLL}$ contains small-signal perturbations in the PLL dynamics ($\omega_{pll}$, $\theta_{pll}$, and internal PLL states), $\Delta x_{PQ}$ corresponds to integrator states of active/reactive power controllers, $\Delta x_{c}$ refers to the integrator states of the current controller, $\Delta x_{LCL}$ encapsulates the filter inductor currents and capacitor voltages in the dq frame, and $\Delta x_{int}$ includes the DC-link voltage dynamics with internal PV--BESS states such as the PV current and battery-side capacitor voltages.

Accordingly, the full state vector of the PV--BESS subsystem becomes $x_{GFL} = [\omega_{pll}, \theta_{pll}, \ \xi_{pll}, \ \xi_P, \xi_Q, \xi_{c,d}, \xi_{c,q}, I_{t,d},\\ I_{t,q}, V_{pcc,d}, V_{pcc,q}, I_{g,d}, I_{g,q}, V_{dc}, i_{pv}, V_{C_b}, V_{C_s}]^T.$ The control input vector for the PV--BESS, denoted $u_{GFL}$, includes the active/reactive power references and local grid voltage at the point of connection, $u_{GFL} = [P^{ref}, Q^{ref}, v_{gd}, v_{gq}]^T$.

The linearized dynamics of the PV--BESS in state-space form are
\begin{equation}
\dot{x}_{GFL} = A_{GFL} x_{GFL} + B_{GFL} u_{GFL},
\end{equation}
where matrices $A_{GFL}$, and $B_{GFL}$ arise from linearization around an operating point. In parallel, the synchronous generator model includes the electromechanical dynamics represented by small-signal deviations in the rotor angle ($\delta_{SG}$), rotor speed ($\omega_{SG}$), $\Delta x_{SG} = [\Delta \delta_{SG}, \Delta \omega_{SG}]^T.$

The full integrated system state vector $x_{sys}$ is then defined as $x_{sys} = [x_{GFL}, x_{SG}]^T.$ The combined input vector $u_{sys}$, encompassing both PV--BESS and SG control inputs, is $u_{sys} = [P^{ref}, Q^{ref}, v_{gd}, v_{gq}, P_{in}]^T.$

Finally, the linearized state-space model of the entire system is expressed as
\begin{equation}
\dot{x}_{sys} = A_{sys} x_{sys} + B_{sys} u_{sys},
\end{equation} 
\begin{equation}
y_{sys} = C_{sys} x_{sys},
\end{equation}
where $A_{sys}$, $B_{sys}$, and $C_{sys}$ characterize the interaction between the GFL-controlled PV--BESS and the synchronous generator, enabling holistic dynamic analysis and control design.
\vspace{-1.5em}
\section{Contingency Modeling in Stochastic Hybrid System Framework}
\label{sec3}

To effectively monitor and manage the dynamic response of distribution grids with IBRs under contingencies, it is essential to develop a framework that captures both their continuous-time dynamics and the discrete events triggered by contingencies. These contingencies may include changes in the system structure, sensor failures, or actuator malfunctions. 
%
When the power system undergoes an abrupt contingency, its dynamics can be modeled as a Randomly Switched Linear System (RSLS), where each discrete operational mode corresponds to a specific scenario of contingency. The system dynamics for a given scenario \(\alpha \in \mathcal{S} = \{1,2,...,n\}\) are described by
\begin{equation}\label{5}
\dot{x}(t) = A(\alpha)x(t) + B(\alpha)u(t), 
\end{equation}
\begin{equation}\label{6}
y(t) = C(\alpha)x(t),  
\end{equation}
for \(t \in [k\tau, (k+1)\tau)\), where \(\tau\) is the total time interval associated with each contingency management cycle. Following \cite{10003976}, each interval is further divided into two subintervals. The first subinterval \(t \in [k\tau, k\tau + \tau_0)\), where \(\tau_0 \ll \tau\), is designated as the detection window, during which real-time measurements \(y(t)\) are compared with a precomputed library of expected responses \(\widehat{y}_\alpha(t)\), corresponding to each possible scenario \(\alpha \in \mathcal{S}\). 
The remaining sub-interval \(t \in (k\tau + \tau_0, (k+1)\tau)\) constitutes the mitigation window, where the best-matching scenario guides state estimation and control implementation until the start of the next detection window. The matrix triplet \((A, B, C)\) remains constant within each interval but may vary across different scenarios \(\alpha\), reflecting structural changes due to contingencies \cite{Stop24}. Thus,
variations in the system's transfer function serve as early indicators of such contingencies. This approach enables the detection of unobservable contingencies—those that cannot be identified using conventional sensing and measurement techniques.
%
%
In this study, we categorize the contingencies into two major types:

\textbf{1. Physical Contingencies:} These include line trips, generator faults, communication disruptions, or converter failures and thus result in changes in the matrices \(A(\alpha)\) and \(B(\alpha)\). 
Mathematically, these can be expressed using indicator functions:
\begin{align}
A(\alpha) &= \sum_{i=1}^n A^{(i)}\, \mathbf{1}_{\{\alpha = i\}}, \\
B(\alpha) &= \sum_{i=1}^n B^{(i)}\, \mathbf{1}_{\{\alpha = i\}}, 
\end{align}
where \(\mathbf{1}_{\{\alpha = i\}}\) equals 1 when \(\alpha = i\), and 0 otherwise.

\textbf{2. Measurement Contingencies:} These affect only the output measurement matrix \(C(\alpha)\), such as sensor failures or spoofing attacks, while the system dynamics remain unchanged. This leads to:
\begin{equation}
C(\alpha) = \sum_{i=1}^m C^{(i)}\, \mathbf{1}_{\{\alpha = i\}}.
\end{equation}

To assess the observability of the system under different scenarios, we use the observability matrix:
\begin{equation}
\mathcal{O}(\alpha) = \begin{bmatrix}
C(\alpha)^T & 
(C(\alpha)A(\alpha))^T & 
\cdots & 
(C(\alpha)A(\alpha)^{n-1})^T
\end{bmatrix}^T.
\end{equation}

Since multiple contingencies may yield systems with similar structures and overlapping eigenvalues, it becomes necessary to design inputs that help distinguish between these scenarios \cite{YUAN202410141}. One such input is the Mode-Modulated Input (MoMI), defined as:
\begin{equation}
U(s) = \frac{a(s)}{b(s)},
\end{equation}
where \(a(s)\) and \(b(s)\) are coprime polynomials, and \(b(s)\) must have at least one root that is not shared with the poles of \(A(\alpha)\) under any scenario. This design ensures that the resulting system outputs remain distinguishable across different contingency scenarios due to non-overlapping modal responses.

\section{Time Series Learning for Contingency Detection within the SHS Framework}
\label{Sec4}
Recent advances in deep learning have demonstrated the effectiveness of attention-based architectures for modeling complex temporal dependencies in sequential data \cite{10.1145/3447548.3467401, app12168085,shimillas2025transformer,katrompas2022recurrence, liu2024time}. Among these, the Transformer has emerged as a powerful and flexible model, originally developed for natural language processing but now widely adopted in time series forecasting, anomaly detection, and classification tasks \cite{shimillas2025transformer,katrompas2022recurrence}. Unlike recurrent neural networks (RNNs), which process sequences sequentially, Transformers operate on the entire sequence in parallel through a self-attention mechanism\cite{REZA2022117275}. This allows the model to efficiently capture both short-term fluctuations and long-range dependencies. 

\subsection{Transformer Model Overview}

Let the input multivariate time series be denoted by \(\mathbf{Z} \in \mathbb{R}^{S \times M}\), where \(S\) is the number of time steps in the detection window, and \(M\) represents the total number of features. In this study, \(\mathbf{Z}\) is constructed by concatenating the system states \(\mathbf{x}(t)\) and outputs \(\mathbf{y}(t)\) from the SHS model defined in ~\eqref{5} and \eqref{6}, such that at each time step \(t\), $\mathbf{z}_t = \begin{bmatrix} \mathbf{x}(t)^\top & \mathbf{y}(t)^\top \end{bmatrix}^\top$,
and the full input sequence is
$\mathbf{Z} = \begin{bmatrix} \mathbf{z}_1^\top & \mathbf{z}_2^\top & \cdots & \mathbf{z}_S^\top \end{bmatrix}^\top$.

\begin{figure}[t!]
  \includegraphics[width=0.55\linewidth,clip]{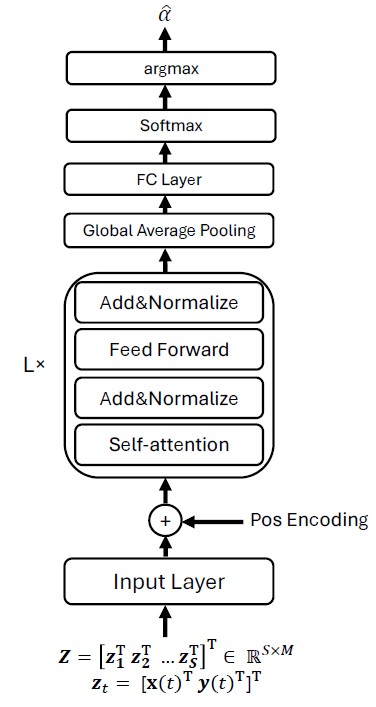}
  \centering
  \captionsetup{width=\linewidth}
  
  \caption{Transformer-based model using multivariate time series input.}
  \label{Trans}
  \vspace{-2em}
\end{figure}

As illustrated in Fig.~\ref{Trans}, each time step input vector \(\mathbf{z}_t \in \mathbb{R}^M\) is linearly projected into a \(d\)-dimensional embedding space via a learnable linear transformation. Positional encodings are then added to these embeddings to encode temporal order, resulting in the input to the Transformer encoder.

The Transformer model consists of \(L\) stacked encoder layers. Each encoder layer contains a multi-head self-attention mechanism followed by a position-wise feedforward network. The multi-head attention enables the model to jointly attend to information from different representation subspaces at different positions, thereby capturing complex temporal dependencies.
Formally, the scaled dot-product attention is computed as:
\begin{equation}
\text{Attention}(\mathbf{Q}, \mathbf{K}, \mathbf{V}) = \text{softmax}\left(\frac{\mathbf{QK}^\top}{\sqrt{d_k}}\right) \mathbf{V},
\end{equation}
where \(\mathbf{Q}\) (queries), \(\mathbf{K}\) (keys), and \(\mathbf{V}\) (values) are matrices derived from the input embeddings through learned linear projections. The parameter \(d_k\) represents the dimensionality of the queries and keys within each attention head, and the scaling by \(\sqrt{d_k}\) stabilizes gradients during training.

Each encoder layer contains \(h\) such attention heads operating in parallel, allowing the model to capture diverse aspects of the input sequence. The outputs of the attention heads are concatenated and passed through another learned linear layer. Residual connections and layer normalization follow both the multi-head attention and feedforward sub-layers to promote stable and efficient training.
After the final encoder layer, global average pooling is applied across the time dimension to obtain a fixed-length latent feature vector \(\bar{\mathbf{h}} \in \mathbb{R}^d\), where \(d\) is the Transformer embedding dimension. This vector summarizes the temporal information from the entire input sequence.

The pooled representation \(\bar{\mathbf{h}}\) is then fed into a fully connected output layer that produces the predicted contingency class probabilities:
\begin{equation}
\hat{{p}} = \text{softmax}(\mathbf{W}_{\text{class}} \bar{\mathbf{h}} + \mathbf{b}_{\text{class}}),
\end{equation}
where \(\hat{{p}} \in \mathbb{R}^{N_c}\) represents the probability distribution over \(N_c\) scenario classes. The predicted scenario class label \(\hat{\alpha}\) is obtained by selecting the class with the highest probability:
\begin{equation}
\hat{\alpha} = \arg\max_{i} \hat{p}_i,
\end{equation}

The model is trained end-to-end using the categorical cross-entropy loss, enabling it to learn temporal patterns associated with different contingency types directly from raw SHS state and output data. This provides a powerful framework for real-time contingency classification. 

\subsection{Integration of SHS with Transformer-Based Learning}
\label{ssec:shs_transformer}
The following describes how we integrate the SHS framework with Transformer-based learning to enhance real-time contingency detection and classification. The integrated approach leverages both the state outputs and network measurements within the SHS representation, facilitating improved classification accuracy. Specifically, within the detection window interval $[k\tau, k\tau + \tau_0)$, we utilize the SHS-modeled system states and network outputs, sampled at a high frequency, as input features for the Transformer model. The sampled multivariate time series for the \(k\)-th detection interval can be represented as
\begin{equation}
\mathbf{Z}^{(k)} = \begin{bmatrix} \mathbf{z}_{k,1} & \mathbf{z}_{k,2} & \dots & \mathbf{z}_{k,S} \end{bmatrix}^\top \in \mathbb{R}^{S \times M},
\end{equation}
where \(S = \frac{\tau_0}{\Delta t}\) denotes the number of samples collected during the detection window, \(\Delta t\) is the sampling interval, and each \(\mathbf{z}_{k,t} \in \mathbb{R}^M\) represents the concatenated system states and output features at time \(t\) within the \(k\)-th detection interval.
This data is processed by the Transformer to classify the system scenario accurately and promptly. 

The overall implementation proceeds in two distinct phases. The first is an offline training phase, in which the Transformer model is trained using historical contingency data and simulations generated from the SHS model. During this phase, multiple contingency scenarios, encompassing both normal operations and various fault conditions, are simulated to generate rich training datasets capturing representative system behaviors. The second phase is the online detection, activated in real-time upon the occurrence of an event. Once trained, the Transformer receives real-time sampled data from the system states and network outputs following a contingency event, classifies the event, and identifies the type of contingency.

\section{SIMULATION}
\label{Sec5}
The performance and effectiveness of the proposed learning-based SHS contingency detection framework is evaluated using the IEEE 33-bus distribution test system, with detailed configuration parameters provided in~\cite{Dolatabadi2020}. The system under consideration consists of generators \(G_1\), \(G_2\), and \(G_4\), each with an active power capacity of \(0.2~\text{MW}\), and one PV--BESS unit with a power capacity of \(0.2~\text{MVA}\). Generator \(G_1\) at bus 1 serves as the slack bus, providing a total power capacity of \(4~\text{MW}\). The system states defined for the dynamic model (outlined in Section~\ref{Sec2}) include the states $x_i = [\delta_i, \omega_i]$ for each generator, where $\delta_i$ and $\omega_i$ represent the rotor angle and rotor speed deviation, respectively, and 16 additional states representing the internal dynamics and control states of the PV–BESS system. Also, two Phasor Measurement Units (PMUs) are installed at generators $G_1$ and $G_2$ to measure rotor angles $\delta_1$ and $\delta_2$.

\begin{figure}[t!]
  \vspace{-0.5em}
  \includegraphics[width=1.05\linewidth,clip]{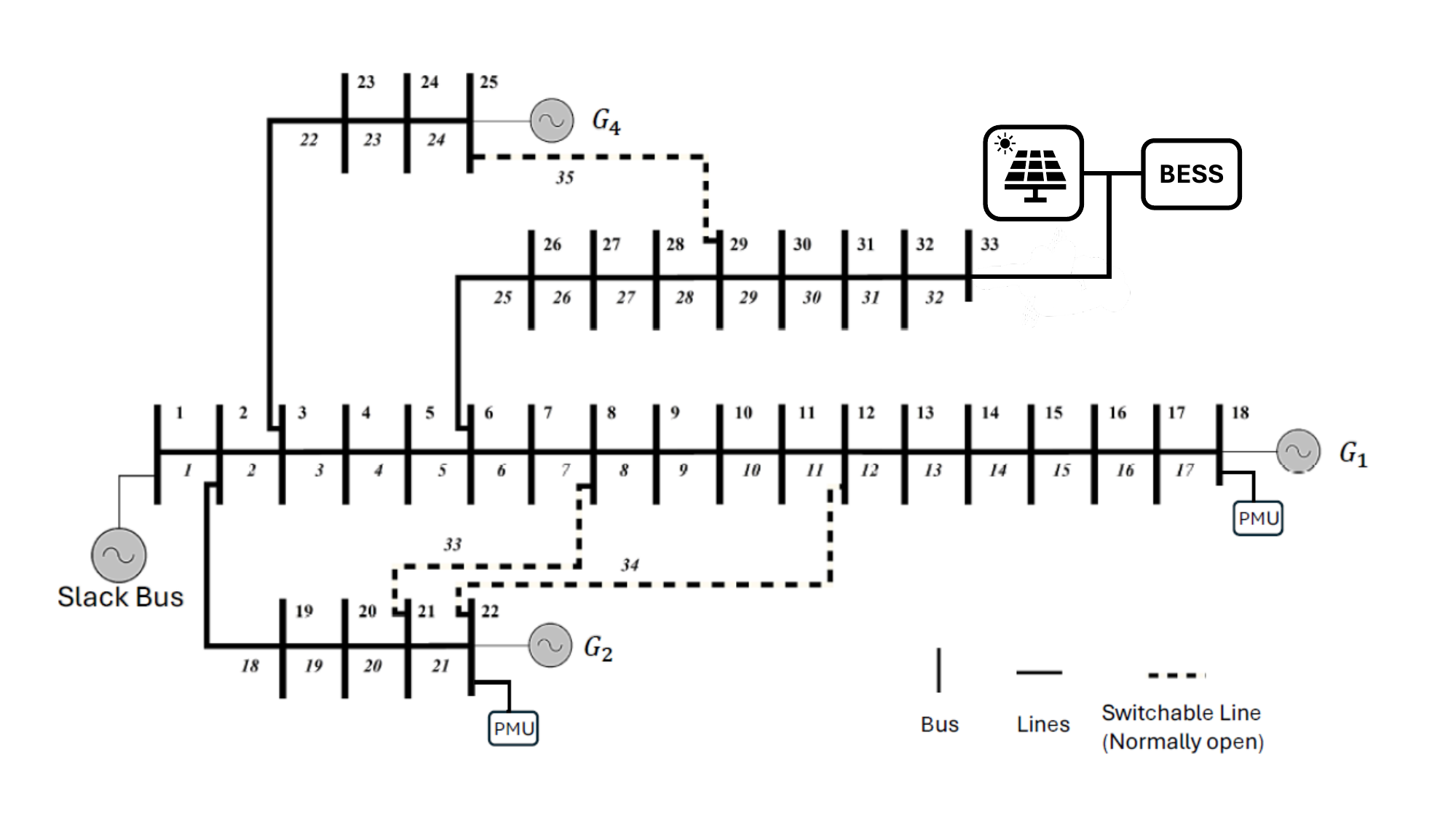}
        \centering
        \captionsetup{width=\linewidth}
        \vspace{-2.5em}
        \caption{Modified IEEE-33 bus systemt \cite{Dolatabadi2020}.}
        \label{33bus}
        \vspace{-1.5em}
\end{figure}

\begin{figure}[t!]
  \includegraphics[width=0.8\linewidth,clip]{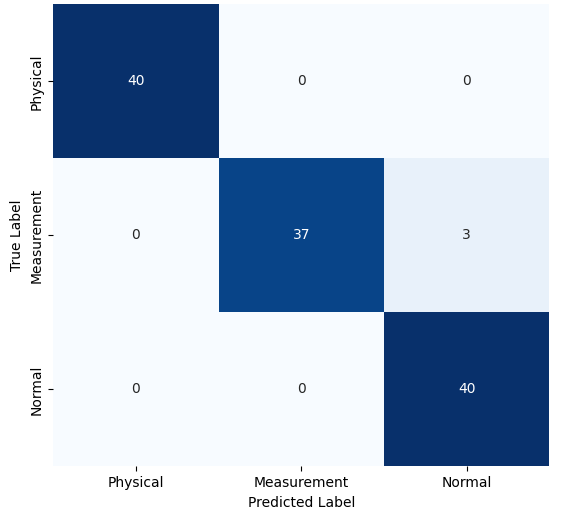}
        \centering
        \captionsetup{width=\linewidth}
        \caption{Performance of the transformer encoder for Training with Accuracy = 98\%.}
        \label{confusion}
        \vspace{-1.5em}
\end{figure}

For training the Transformer model, the dataset is constructed from the states and outputs of the dynamic model as  $\mathbf{Z} = [\delta_1, \delta_2, \delta_4, \omega_1, \omega_2, \omega_4, x_{\text{PV-BESS},1}, \dots, x_{\text{PV-BESS},16}]^\top $, sampled using MATLAB environment. The sampling interval is defined as \( \Delta t = 0.001~\text{s} \), resulting in \( S = \tau_0/\Delta t = 30 \) samples per detection window \( \tau_0 = 0.03~\text{s} \). The dataset comprises $n$ = 600 scenarios, equally distributed among three distinct operational classes: normal operation, physical contingencies, and measurement contingencies, with 200 scenarios per class. 
\begin{figure}[t!]
  
  \includegraphics[width=1.05\linewidth,clip]{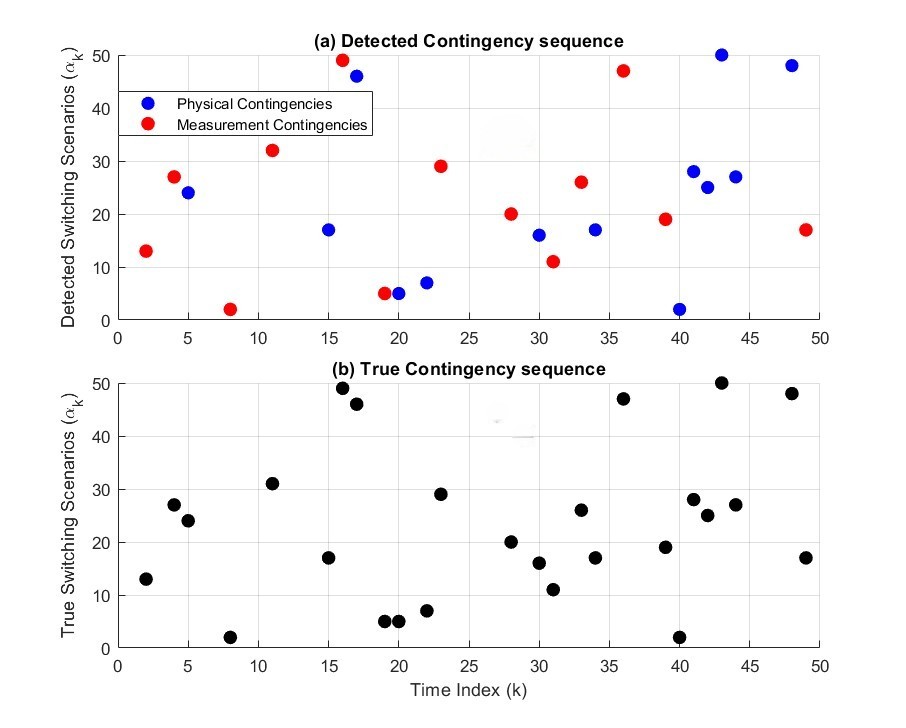}
        \centering
        \captionsetup{width=\linewidth}
        \vspace{-2.5em}
        \caption{Switching sequence of contingencies: (a) Detected scenarios; (b) True scenarios.}
        \label{detection}
        \vspace{-1.5em}
\end{figure}
The dataset was split into 80\% training and 20\% test subsets. To optimize the Transformer model configuration, a grid search was conducted over various hyperparameters. The best-performing model consists of \(L=6\) encoder layers, each utilizing 8 attention heads, a hidden dimension size of 256 in the position-wise feedforward network, an embedding dimension \(d=64\), a dropout rate of 0.1, a learning rate of 0.0001, and a batch size of 16. The model was trained for 100 epochs. The performance of the trained Transformer-based model was evaluated using a confusion matrix, depicted in Fig.~\ref{confusion}. The model achieved an overall accuracy of 98\%, successfully distinguishing physical contingencies and measurement contingencies.

To validate the online applicability of the trained Transformer model, an additional set of $n$ = 200 contingency scenarios was simulated. Fig.~\ref{detection} illustrates the true and detected contingency sequences for the first 50 scenarios to provide a visualization of the model's performance. In the figure, black dots represent the true contingency switching sequence, while the detected sequence is shown using blue and red dots, corresponding to physical contingencies (blue) and measurement contingencies (red), respectively. The physical contingencies correspond to \(N-1\) line outage scenarios, and the measurement contingencies involve sensor values varying from a 50\% decrease to a 50\% increase, in increments of 5\%. The online detection accuracy reached 96\%. 
These results affirm the practical effectiveness of the proposed Transformer-based SHS framework for fast and accurate contingency detection and classification in distribution networks integrating IBRs.

\section{Conclusions}
\label{Sec6}
This paper presented an SHS-based modeling framework for contingency detection and classification in modern distribution grids with integrated GFL-controlled PV–BESS and synchronous generators. The SHS approach effectively captured the continuous dynamics of power systems alongside discrete contingency events, while a Transformer-based time series learning method was introduced for real-time detection. Two primary classes of contingencies—physical and measurement—were accurately identified, with simulation results on the IEEE 33-bus test system achieving up to 98\% offline and 96\% online detection accuracy. The proposed SHS–Transformer framework thus provides a practical and scalable tool for enhancing situational awareness and operational resilience in distribution networks with high penetration of IBRs. Although larger and more complex networks would introduce higher-dimensional inputs and a broader range of contingencies, the Transformer’s parallel self-attention mechanism is well-suited for scaling to such environments. Future work will extend this framework to larger benchmarks and explore dimensionality reduction and distributed training strategies to manage computational demands.

\bibliographystyle{ieeetr}
\bibliography{references}

\end{document}